\documentclass{article}

\usepackage{microtype}
\usepackage{graphicx}
\usepackage{amsfonts}
\usepackage{booktabs} 
\usepackage{etoolbox}
\usepackage{soul}
\usepackage[font={footnotesize}]{subcaption}
\usepackage{hyperref}
\usepackage{nameref}
\usepackage[]{appendix}
\usepackage{stfloats}
\usepackage[bottom]{footmisc}

\usepackage{hyperref}

\usepackage{amsmath}
\usepackage[accepted]{icml2021}

\icmltitlerunning{High Frequency EEG Artifact Detection with Uncertainty via Early Exit Paradigm}

\begin{document}

\newbool{showComments}
\booltrue{showComments}
\ifbool{showComments}{%
\newcommand{\cm}[1]{\sethlcolor{green}\hl{[Cecilia: #1]}}
\newcommand{\pl}[1]{\sethlcolor{orange}\hl{[Pietro: #1]}}
\newcommand{\lo}[1]{\sethlcolor{yellow}\hl{[Lorena: #1]}}
\newcommand{\ac}[1]{\sethlcolor{cyan}\hl{[Alex: #1]}}
}{
\newcommand{\cm}[1]{}
\newcommand{\jc}[1]{}
\newcommand{\lo}[1]{}
\newcommand{\gr}[1]{}
}
\newcommand{\e}{EEG }
\newcommand{\us}{E$^4$G }

\twocolumn[
\icmltitle{High Frequency EEG Artifact Detection with Uncertainty\\ via Early Exit Paradigm}



\icmlsetsymbol{equal}{*}
\vspace{-0.12in}

\begin{icmlauthorlist}
\icmlauthor{Lorena Qendro}{equal,uc}
\icmlauthor{Alexander Campbell}{equal,uc,ati}
\icmlauthor{Pietro Liò}{uc}
\icmlauthor{Cecilia Mascolo}{uc}
\end{icmlauthorlist}

\icmlaffiliation{uc}{University of Cambridge}
\icmlaffiliation{ati}{The Alan Turing Institute}


\icmlkeywords{EEG, Artifact detection, Time series, Uncertainty, Deep Learning}

\vskip 0.3in
]



\printAffiliationsAndNotice{\icmlEqualContribution} 

\begin{abstract}
Electroencephalography (EEG) is crucial for the monitoring and diagnosis of brain disorders. However, EEG signals suffer from perturbations caused by non-cerebral artifacts limiting their efficacy. Current artifact detection pipelines are resource-hungry and rely heavily on hand-crafted features. Moreover, these pipelines are deterministic in nature, making them unable to capture predictive uncertainty. We propose E$^4$G , a deep learning framework for high frequency EEG artifact detection. Our framework exploits the early exit paradigm, building an implicit ensemble of models capable of capturing uncertainty. We evaluate our approach on the Temple University Hospital EEG Artifact Corpus (v2.0) achieving state-of-the-art classification results. In addition, E$^4$G provides well-calibrated uncertainty metrics comparable to sampling techniques like Monte Carlo dropout in just a single forward pass.  E$^4$G opens the door to uncertainty-aware artifact detection supporting clinicians-in-the-loop frameworks.
\end{abstract}
\section{Introduction}
Electroencephalography (\e) is a non-invasive imaging technique for measuring electrical activity of the brain. \e is widely used for monitoring and diagnosing brain disorders ranging from epilepsy~\citep{acharya2013automated} and tumors to depression~\citep{de2019depression} and sleep disorders~\citep{perslev2019u}. An important preprocessing step in analyzing EEG signals is the removal of artifacts: waveforms caused by external or physiological factors that are not of cerebral origin. \e artifacts are often mistaken for seizures due to their morphological similarity in amplitude and frequency, leading to increased rates of false alarms~\citep{ochal2020temple}. As such, the removal of artifacts is critical to the use of EEG in clinical practice. To date, the majority of existing deep learning approaches for \e artifact detection still use hand-crafted features~\cite{leite2018deep, lee2020eeg} and fail to provide uncertainty estimations, limiting the trustworthiness of their predictions. 

We propose \textbf{E}arly \textbf{E}xit \textbf{EEG} (\us), a general deep learning framework for end-to-end per time point \e artifact detection with uncertainty. We demonstrate that integrating the early exit (EE) paradigm~\cite{huang2017multi,montanari2020eperceptive} into state-of-the-art time series segmentation models~\cite{ronneberger2015u, perslev2021u} allows for robust and uncertainty-aware predictions. Early exit points from any deep learning architecture form an implicit ensemble of models from which predictive uncertainty can be estimated. In contrast to sampling-based Bayesian approaches like Monte Carlo dropout~\cite{gal2016dropout}, \us can provide well calibrated uncertainty that is better or comparable to MCDrop, especially for incorrect predictions as measured by predictive entropy (0.45 vs 0.53) and Brier score (0.53 vs 0.55), in a single forward pass.

Our approach can aid clinician-in-the-loop frameworks, where the majority of artifacts are automatically detected with high confidence and the uncertain ones sent to the human expert for further investigation.

\section{Related work}

Existing methods for \e artifact detection and removal tend to rely on signal filtering \cite{mowla2015artifacts, seifzadeh2014comparison} and/or blind source separation techniques such as independent component analysis (ICA) \cite{winkler2014robust, hazra2010modified}. Both approaches require expert domain knowledge, are time consuming, and computationally expensive~\cite{jafari2017eeg, castellanos2006recovering}. Deep learning approaches have been successfully applied to the task of artifact detection~\cite{hu2015removal, leite2018deep, khatwani2021flexible, cisotto2020comparison}. The reliance on hand-crafted features and window-based classification limits their scalability and generalisability. The work of~\citet{perslev2021u} showcase an efficient fully convolutional architecture for the task of \e sleep stage classification. Unlike \us, all these methods do not perform per time point predictions and are deterministic thereby failing to provide predictive uncertainty.
\section{Method}

Let $\mathbf{x}_{1:T}\in\mathbb{R}^{M \times T}$ denote a time series with $M$ channels of length $T$, while $\mathbf{y}_{1:T} \in \{0, 1\}^{C \times T}$ is a corresponding time series of class labels. An artifact is defined as a sub-sequence of time points $(\mathbf{y}_t, \mathbf{y}_{t+s}) \subseteq \mathbf{y}_{1:T}$ of length $s$ such that $1 < s \leq T$.

\subsection{Ensemble of early exits}

Let $f_\theta(\cdot)$ represent any multi-layered neural network architecture with parameters $\theta$ which can be decomposed into $B$ blocks (or layers)
\begin{equation}
f_\theta(\mathbf{x}_{1:T}) = (f^{(B)} \circ f^{(B-1)} \circ \dots \circ f^{(1)})(\mathbf{x}_{1:T})
\end{equation}
where $\theta = \cup_{i=1}^B \theta_i $ and  $\circ$ denotes function composition  $(f^{(i)} \circ f^{(j)})(\cdot) = f_{\theta_i}(f_{\theta_j}(\cdot))$ when $i \neq j$.

Given the intermediary output $\mathbf{h}^{(i)} = f_{\theta_i}(\mathbf{h}^{{(i-1)}})$ of the $i$-th block, let $g_{\phi_i}^{(i)}(\cdot)$ denote a corresponding exit branch with parameters $\phi_i$. Each exit branch maps the intermediary output to a prediction $\hat{\mathbf{y}}_{1:T}^{(i)} = g_{\phi_i}(\mathbf{h}^{(i)})$. The set of early exit predictions
\begin{equation}
\label{ensebmle}
f_\theta(\mathbf{x}_{1:T}) = \{ \hat{\mathbf{y}}^{(1)}_{1:T}, \dots, \hat{\mathbf{y}}^{(B)}_{1:T} \},
\end{equation}
where each $\hat{\mathbf{y}}^{(i)}_{1:T}$ represents the per time point prediction of the $i$-th model, constituting an implicit ensemble of networks.

The typical use case of EEs is for conditional computation models where each exit is gated on a satisfied criterion such as accuracy in order to save time and computation~\cite{bolukbasi2017adaptive}. However, we exploit this paradigm to have \emph{more than one} prediction at training \emph{and} inference time (Equation~\ref{ensebmle}), equipping the model with awareness of uncertainty which can be leveraged for a variety of tasks ranging from robustness (out-of-distribution detection~\cite{ovadia2019can}) to decision making (optimal group voting strategies~\cite{shahzad2013comparative}).

\subsection{Training with early exists}

\begin{figure}[ht]
\begin{center}
\includegraphics[width=0.8\columnwidth]{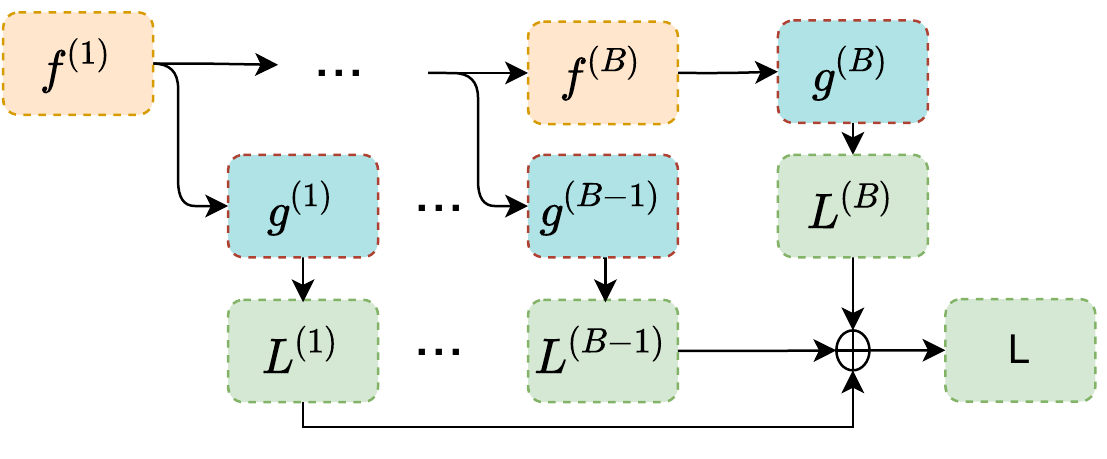}
    \caption{Training with early exits.  Each exit branch $g^{(i)}$ propagates the prediction error in relation to the ground truth label $L^{(i)}$, training the blocks that precede each exit point $f^{(i-1)}$ for $i=1,\dots,B$.}
    \label{fig:ee_train}
\end{center}
\vspace{-0.12in}
\end{figure}

To train a model with EEs, the loss function is a composition of the individual predictive losses of each exit (see Figure~\ref{fig:ee_train}). As such, each prediction propagates the error in relation to the ground truth label to the blocks preceding that exit. This can be represented as a weighted sum
\begin{equation}
L = \sum_{i=1}^B \alpha_i L^{(i)}(\mathbf{y}_{1:T}, \hat{\mathbf{y}}_{1:T}^{(i)} )
\end{equation}
where  $L^{(i)}(\cdot, \cdot)$ is the loss function for the $i$-th block's exit and $\alpha_i \in \{0, 1\}$ is a weight parameter corresponding to the relative importance of the exit. Increasing the weight at a specific exit point forces the network to learn better features at the preceding layers~\cite{scardapane2020should}.

This procedure allows for an efficient training of the whole ensemble in one go. Furthermore, differently from traditional deep ensembles using varying architectures, EEs as an ensemble allow for the earlier exits to incorporate feature learnt from the exits that follow.

\subsection{Inference and predictive uncertainty}

During inference, a single forward pass of the network produces a sequence of predictions, each \emph{potentially} more accurate than the previous one. The overall prediction of the ensemble can be computed as the mean of a categorical distribution obtained from aggregating the predictions from the individual exits

\begin{equation}
p(\mathbf{y}_{1:T}| \mathbf{x}_{1:T})\approx \frac{1}{B} \sum_{i=1}^B \text{Softmax}(\hat{\mathbf{y}}^{(i)}_{1:T})
\label{confidence}
\end{equation}

where the softmax function is applied time pointwise. Although the softmax function scales inputs to the range $[0, 1]$, it is not a valid measure of probability since it is based on a point estimate. In contrast, as shown in Equation~\ref{confidence} (Appendix~\ref{appendices}), using our framework, real probability can be approximated from $B$ different possible outcomes from a single underlying data generating process.

Given the approximate distribution in Equation~\ref{confidence} (Appendix~\ref{appendices}), \us can therefore capture predictive uncertainty as measured by metrics such as entropy:
\begin{equation}
 H(\mathbf{y}_{1:T}| \mathbf{x}_{1:T}) = - \sum_{\mathbf{y}_{1:T} \in \{0, 1\}} p(\mathbf{y}_{1:T}| \mathbf{x}_{1:T}) \log p(\mathbf{y}_{1:T}| \mathbf{x}_{1:T}). 
\label{entropy}
\end{equation}

Predictive uncertainty metrics give an interpretation of model decisions that can inform post-processing heuristics which would allow for the isolation highly uncertain predictions for further investigation by a human expert, thereby, increasing the overall trust and accuracy of the model.
\section{Experiments}

\subsection{Dataset}

We evaluate our framework on the Temple University \e Artifact Corpus (v2.0) (TUH-A) ~\cite{hamid2020temple}. The dataset is composed of real EEG signals from 213 patients contaminated with artifacts which are human expert annotated. The artifacts range from chewing, eye movement, muscular movement, shivering and electrode errors. To best of our knowledge, we are the first to use this dataset in an automatic artifact detection framework. Appendix~\ref{dataset} contains further details on the dataset and preprocessing.

\subsection{Implementation}

\begin{figure}[!hb]
\begin{center}
\includegraphics[trim=1 1 0 0,clip,width=1.0\columnwidth]{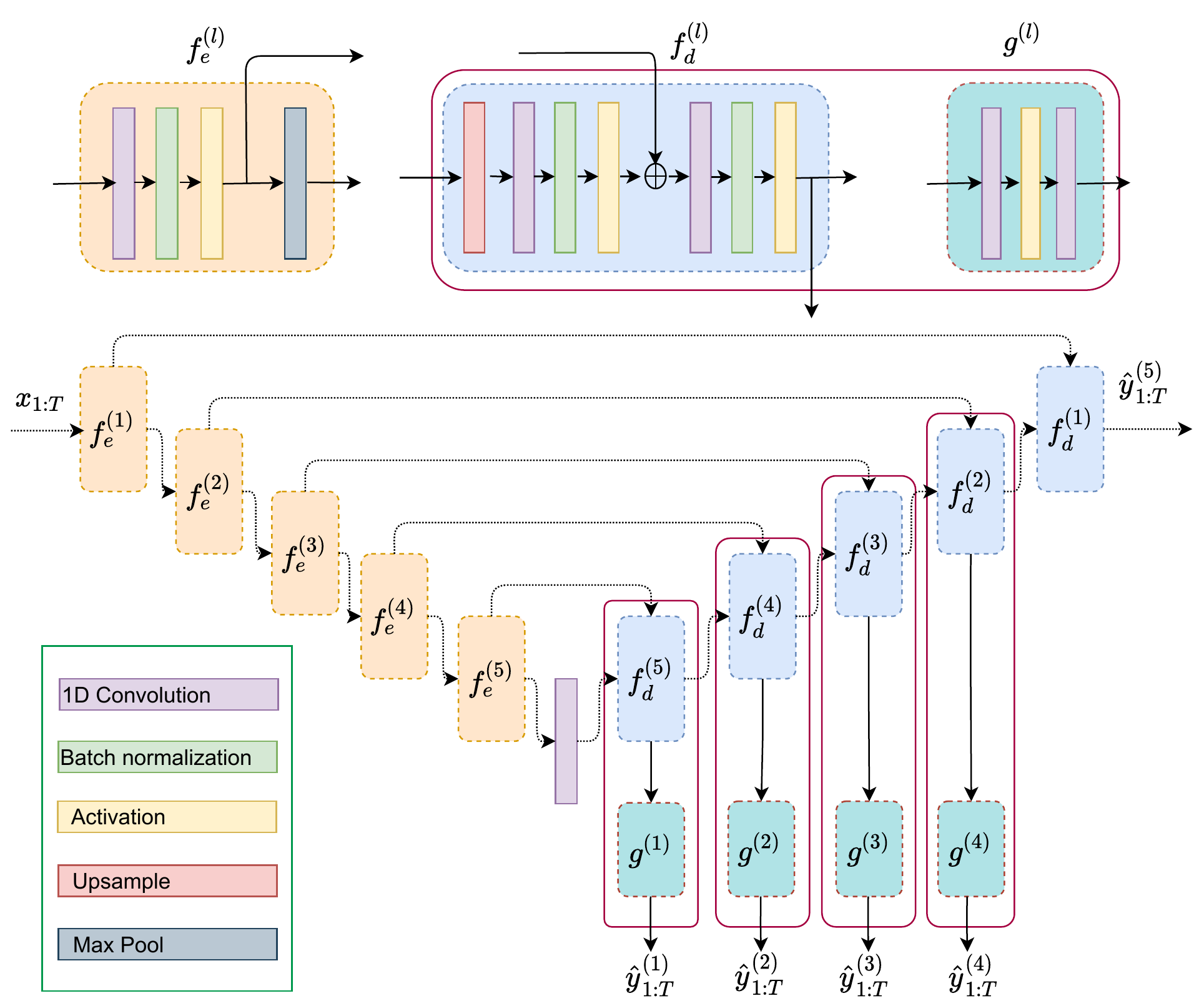}
    \caption{Overview of \us based on a U-Net. The neural networks $f_e$, $f_d$, and $g$ represent the encoder, decoder and exit blocks, respectively (top). Each block in the encoder-decoder is connected by skip connections which perform feature concatenation (bottom). Exit blocks are placed after each hidden decoder block.}
    \label{fig:overview_eeg}
\end{center}
\end{figure}

For \us, the backbone neural network $f_\theta(\cdot)$ is implemented as a temporal U-Net~\cite{perslev2021u}, a fully convolutional encoder-decoder architecture for time series segmentation. We place exit blocks $g_{\phi_i}(\cdot)$ after each decoder block resulting in 4 predictions plus the output layer making a ensemble of $B=5$ as depicted in Figure~\ref{fig:overview_eeg}. The ensemble size is in line with previous work suggesting the optimal number of samples needed for well calibrated uncertainty~\cite{ovadia2019can, qendro2021benefit, qendro2021stochastic}. 

We compare \us with two baselines both using the same backbone architecture. The first is a vanilla U-Net with no early exits and the second adds dropout at the end of each decoder block with a probability of $p=0.2$. The latter model is used for Monte Carlo dropout (MCDrop)~\cite{gal2015bayesian} based on 5 forward passes during testing.

 For training, we use a joint loss composed of an equally weighted cross entropy loss and dice loss~\cite{taghanaki2019combo, isensee2019nnu}. All exits are considered as equally important by setting $\alpha_i =1$. We train all models for a maximum of 200 epochs with a batch size of 50 using the Adam optimizer~\cite{kingma2014adam} with a learning rate of 1$e$-3.  To prevent overfitting, model training was stopping based on no improvement in validation F1 score. All experiments are performed in PyTorch~\cite{paszke2019pytorch}. More details about model implementation can be found in Table~\ref{implementation_table} in Appendix~\ref{implementation}.
 
\begin{table*}[!tp]
\centering
\huge
\resizebox{\textwidth}{!}{
\begin{tabular}{cccccccccc}
\toprule
\multirow{2}{*}{\textbf{Model}} & \multirow{2}{*}{\textbf{F1 Score $\uparrow$}} & \multirow{2}{*}{\textbf{Precision $\uparrow$}} & \multirow{2}{*}{\textbf{Recall $\uparrow$}} & \multicolumn{2}{c}{\textbf{Predictive Entropy}} & \multicolumn{2}{c}{\textbf{Brier Score}} & \multicolumn{2}{c}{\textbf{Predictive Confidence}} \\ \cmidrule(l){5-10} 
\multicolumn{4}{c}{\textbf{}} & \textbf{True $\downarrow$} & \textbf{False $\uparrow$} & \textbf{True $\downarrow$} & \textbf{False $\uparrow$} & \textbf{True $\uparrow$} & \textbf{False $\downarrow$} \\ \midrule
Vanilla U-Net & 0.840$\pm$.06 & 0.854$\pm$.03 & 0.832$\pm$.11 & -- & -- & -- & -- & -- & -- \\
MCDrop U-Net & 0.813$\pm$.07 & 0.821$\pm$.04 & 0.807$\pm$.10 & 0.243$\pm$.01 & 0.457$\pm$.01 & 0.023$\pm$.00 & 0.528$\pm$.01 & 0.905$\pm$.00 & 0.222$\pm$.01  \\
Early Exit EEG (\us) & 0.838$\pm$.06 & 0.853$\pm$.03 & 0.829$\pm$.11 & 0.278$\pm$.01 & 0.531$\pm$.01 & 0.030$\pm$.001 & 0.547$\pm$.01 & 0.886$\pm$.01 & 0.274$\pm$.01 \\ \bottomrule
\end{tabular}
}
\caption{Classification results and uncertainty metrics for EEG artifact detection on TUH-A. Reported results are the mean and standard deviation across 5 runs, each with a different random seed. The best results correspond to low uncertainty for true (right) predictions ($\downarrow$) and high uncertainty for false (wrong) predictions ($\uparrow$).}
\label{tab:metrics}
\end{table*}

\subsection{Results}

Table~\ref{tab:metrics} presents results for the task of per time point EEG artifact detection on the TUH-A test dataset. In addition to standard classification metrics, we consider Brier score, predictive confidence and predictive entropy as measures of uncertainty (see Appendix~\ref{uncertainty_metrics} for a detailed description). 

From the results it is clear that training with EEs does not degrade F1 score compared to the vanilla baseline (83.8\% vs 84\%). Instead, activating dropout during inference in MCDrop lowers F1 score by $~3\%$, suggesting a higher number of samples is required at test time to compensate for the dropped weights during inference. Already, by considering only 5 samples MCDrop increases latency by 5.2x compared to the vanilla baseline, while \us increases latency only marginally (by only 1.1x).~\footnote{Latency is calculated based on the average of 5 runs on the whole test set.}

Importantly, \us can provide well calibrated uncertainty that is better or comparable to MCDrop, especially for incorrect predictions as measured by predictive entropy (0.45 vs 0.53) and Brier score (0.53 vs 0.55), without incurring the computation overhead of the latter framework. Particularly within a medical setting, a highly uncertain false prediction, like the one provided by \us, is more informative for a clinician-in-the-loop framework, where uncertain samples could be transferred to a human expert for further investigation~\cite{leibig2017leveraging}.

\begin{figure*}
\centering
\begin{subfigure}[b]{0.475\textwidth}
\centering
\includegraphics[width=\textwidth]{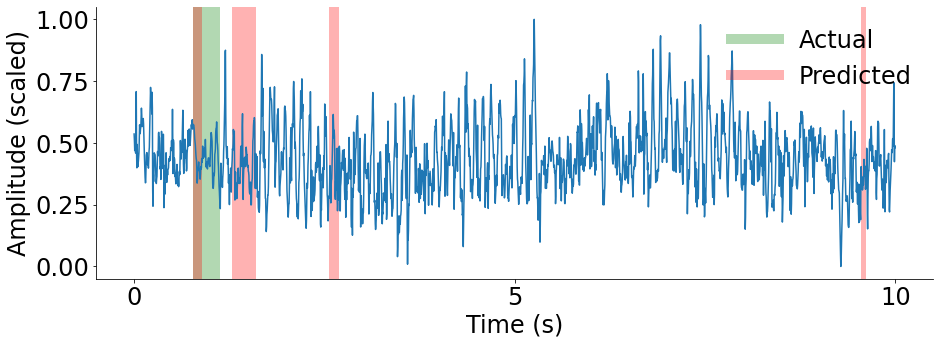}
\caption[]%
{{\small Eye movement, exit 1}}    
\label{fig:eye movement exit 1}
\end{subfigure}
\hfill
\begin{subfigure}[b]{0.475\textwidth}  
\centering 
\includegraphics[width=\textwidth]{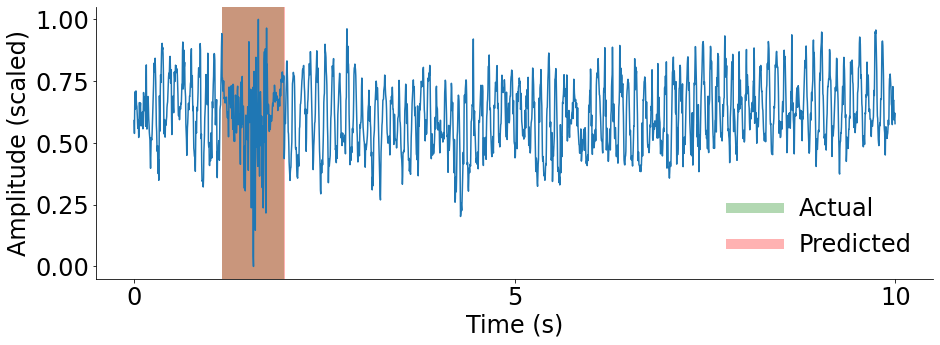}
\caption[]%
{{\small Muscle movement, exit 1}}    
\label{fig:muscle movement exit 1}
\end{subfigure}
\vskip\baselineskip
\begin{subfigure}[b]{0.475\textwidth}   
\centering 
\includegraphics[width=\textwidth]{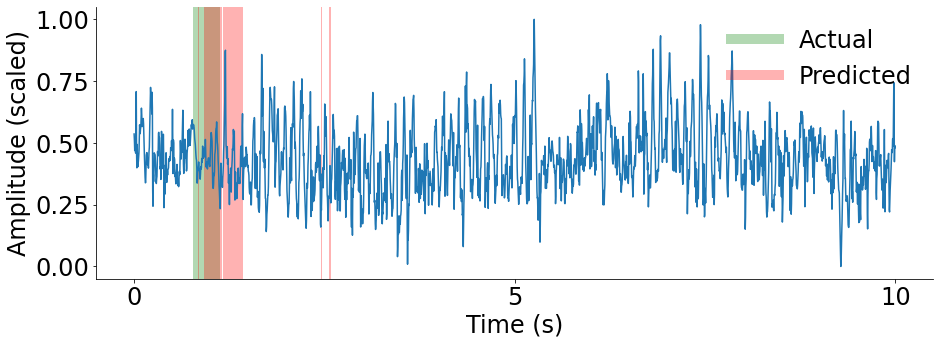} 
\caption[]%
{{\small Eye movement, exit 5}}    
\label{fig:eye movement exit 5}
\end{subfigure}
\hfill
\begin{subfigure}[b]{0.475\textwidth}   
\centering 
\includegraphics[width=\textwidth]{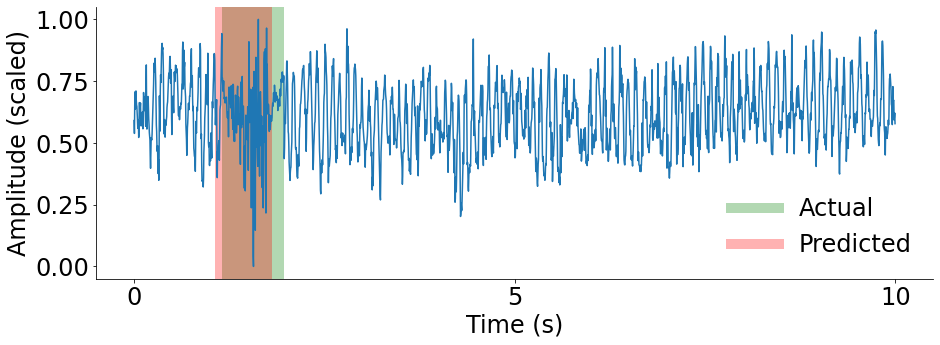}
\caption[]%
{{\small Muscle movement, exit 5}}    
\label{fig:muscle movement exit 5}
\end{subfigure}
\caption[]
{\small Per time point \e artifact predictions for Exits 1 and 5 within a 10 second window. Figure (a) and (c) show disagreement on how the artifact is distributed across time. Figure (b) and (d) show exit disagreement on the start and end point of the artifact. [Green = Actual, Red = Predicted, Brown = True prediction (Red + Green)]} 
\label{fig:chewing and muscle movement exits}
\end{figure*}

\subsection{Recommendations for using predictive uncertainty}

Analyzing the results of the individual exits from \us allows for uncertainty-informed decision making. Figure~\ref{fig:eye movement exit 1} and Figure~\ref{fig:eye movement exit 5} show an example for an eye movement artifact on a 10 second window. The two exits clearly disagree since their predictions (red sections) highlight the presence of the artifact in contradicting areas, particularly for the time points furthest away from the true occurrence of the artifact. Given high uncertainty of a false prediction, an automatic recommendation can be made for an expert clinician to review the sample~\cite{garcia2020uncertainty, xia2021uncertainty}.

On the other hand, Figure~\ref{fig:muscle movement exit 1} and~\ref{fig:muscle movement exit 5} show an example where the two exits agree over when the majority of a muscle movement artifact occur (brown sections). However, uncertainty is still apparent over the start and end time of the artifact. This is valuable information that can be used to automatically detect artifact boundaries for downstream removal and interpolation~\cite{yang2018automatic}.

Although we only show two of the exits, Table~\ref{artifact} in Appendix~\ref{model_predictions} shows results for all exits in the ensemble as well as examples of more artifacts. The strength of our framework is in exploiting exit predictive diversity for better performance and interpretability through the use of different ensemble aggregating strategies such as majority voting, model averaging and automatic thresholding which we leave for future work.
\section{Conclusion}

We introduce a general framework for enabling uncertainty quantification in any feed-forward deep neural network via the EE paradigm. Using the largest publicly available \e artifact dataset, we evaluate our approach on the task of artifact detection and demonstrate how uncertainty, via \us,  can be used to inform decision making. Compared to the commonly used method of uncertainty quantification, Monte Carlo dropout (MCDrop), \us performs better in terms of F1 score while providing comparable well-calibrated uncertainties in a single forward pass. In addition, unlike MCDrop our framework has access to uncertainty within the ensemble during training. Other methods that incorporate this functionality tend only to be pure Bayesian approaches that struggle with underfitting at scale and parameter efficiency~\cite{dusenberry2020efficient}. Future work will include applying the EE paradigm to other architectures as well as other medical datasets like electronic health records. We envision extending our framework to exploit the uncertainty during training for adaptive decision making on joint tasks such as interpolation.

\section{Acknowledgments}
This work is supported by Nokia Bell Labs through their donation for the Centre of Mobile, Wearable Systems and Augmented Intelligence, ERC Project 833296 (EAR), as well as The Alan Turing Institute under the EPSRC grant EP/N510129/1.

\nocite{langley00}

\bibliography{bibliography}
\bibliographystyle{icml2021}
\clearpage
\newpage

\appendix
\begin{appendices}
\label{appendices}

\section{Dataset}
\label{dataset}
From the TUH-A dataset, a common set of 21 \e channels were retained from all patients and signals were resampled to the majority sampling frequency of $250$ Hz. Furthermore, all EEG signals were bandpass filtered (0.3-40 Hz) using a second-degree Butterworth filter and notch filtered at the power lower frequency 60 Hz similar to previous work~\cite{svantesson2021virtual}. Finally, segments of clean and artifact samples of length 10s were created based on an analysis of the duration of the majority of artifacts. We treat each channel independently resulting in each sample $\mathbf{x}_{1:T}$ having size $M$=1 and $T$=2,500.

Since the dataset's artifact classes are highly imbalanced, we augment the data by performing temporal window shifting and signal mixing~\citep{cheng2020subject, chiu2020mixing} making sure not to mix signals from different \e channels and patients. From the augmented dataset of 24,000 samples, we construct masks over time which act as a per time point labels specifying either the presence of an artifact, $y_t=1$ or clean signal $y_t=0$. Finally, we split the data into 80\%, training, 10\% validation and 10\% testing keeping patient independent between datasets. 

\section{Implementation}
\label{implementation}

The core model was based on a temporal U-Net suggested by~\citet{perslev2021u}. Table~\ref{implementation_table} contains model implementation details.

\begin{table}[h]
\Huge 
\resizebox{0.49\textwidth}{!}{
\begin{tabular}{@{}llllll@{}}
\toprule
\multicolumn{2}{c}{\textbf{Encoder Block}} & \multicolumn{2}{c}{\textbf{Decoder Block}} & \multicolumn{2}{c}{\textbf{Exit Block}} \\ \midrule
\textbf{Layer}       & \textbf{Output}     & \textbf{Layer}       & \textbf{Output}     & \textbf{Layer}     & \textbf{Output}   \\ \midrule 
Conv1D, BN, ELU      & 5 x 2500            & Upsample             & 22 x 156            & Upsample           & 16 x 2500          \\
Max Pool             & 5 x 1250            & Conv1D, BN, ELU      & 16 x 156            & Conv1D             & 2 x 2500           \\
                     &                     & Concatenate          & 32 x 156            &                    &                    \\
                     &                     & Conv1D, BN, ELU      & 16 x 156            &                    &                    \\ \midrule 
Conv1D, BN, ELU      & 7 x 1250            & Upsample             & 16 x 312            & Upsample           & 12 x 2500          \\ 
Max Pool             & 7 x 625             & Conv1D, BN, ELU      & 12 x 312            & Conv1D             & 2 x 2500           \\
                     &                     & Concatenate          & 24 x 312            &                    &                    \\
                     &                     & Conv1D, BN, ELU      & 12 x312             &                    &                    \\  \midrule 
Conv1D, BN, ELU      & 9 x 625             & Upsample             & 12 x 625            & Upsample           & 9 x 2500           \\
Max Pool             & 9 x 312             & Conv1D, BN, ELU      & 9 x 625             & Conv1D             & 2 x 2500           \\
                     &                     & Concatenate          & 18 x 625            &                    &                    \\
                     &                     & Conv1D, BN, ELU      & 9 x 625             &                    &                    \\  \midrule 
Conv1D, BN, ELU      & 12 x 312            & Upsample             & 9 x 1250            & Upsample           & 7 x 2500           \\   
Max Pool             & 12 x 156            & Conv1D, BN, ELU      & 7 x 1250            & Conv1D             & 2 x 2500           \\
                     &                     & Concatenate          & 14 x1250            &                    &                    \\
                     &                     & Conv1D, BN, ELU      & 7 x 1250            &                    &                    \\  \midrule 
Conv1D, BN, ELU      & 16 x 156            & Upsample             & 7 x 2500            &                    &                    \\
Max Pool             & 16 x 78             & Conv1D, BN, ELU      & 5 x 2500            &                    &                    \\
                     &                     & Concatenate          & 10 x 2500           &                    &                    \\
                     &                     & Conv1D, BN, ELU      & 5 x 2500            &                    &                    \\  \midrule 
Conv1D, BN, ELU      & 22 x 78             & Conv1D, ELU          & 5 x 2500            &                    &                    \\
                     &                     & Conv1D               & 2 x 2500            &                    &                    \\ \bottomrule
\end{tabular}
}
\caption{Implementation details of \us. Shapes are based on an input size of $M=1$ and $T=2500$. [Max Pool = Max pooling with kernel size 2 and stride 1, Upsample = nearest-neighbor interpolation, Conv1D = 1D convolution with kernel size 4, stride 1, BN = Batch normalization, ELU = ELU activation function].
}
\label{implementation_table}
\end{table}

\section{Results}

\subsection{Uncertainty metrics}
\label{uncertainty_metrics}

Let $\mathbf{y}_{1:T} = \{\mathbf{y}_1, \dots, \mathbf{y}_T\}$ where each $\mathbf{y}_t \in \{0, 1\}^C $ denote the true per time point class labels and $\mathbf{\hat{y}}_{1:T}$ where each $\mathbf{\hat{y}}_{t} \in \mathbb{R}^C$ denote the predicted per class and time point logits.

\paragraph{Brier score} is a measure of the accuracy of predicted probabilities. The Brier score for a single sample is defined
\begin{equation}
    BS = \frac{1}{T} \sum_{t=1}^T (\bar{\mathbf{y}}_t - \mathbf{y}_{t})^2\\
\end{equation}
where $\bar{\mathbf{y}}_t = \frac{1}{B}\sum_{i=1}^B\text{Softmax}(\mathbf{\hat{y}}^{(i)}_{t})$ is the average vector of class probabilities over all exits for a given time point.

\paragraph{Predictive confidence} measures the largest predicted class probability. Predictive confidence is defined as
\begin{equation}
PC =  \frac{1}{T} \sum_{t=1}^T \text{max} (\bar{\mathbf{y}}_t).
\label{confidence}
\end{equation}

\paragraph{Predictive entropy} measure the average amount of information in the predicted distribution. Predictive entropy is defined in Equation~\ref{entropy}.

\subsection{Model predictions}
\label{model_predictions}

\begin{figure*}[htbp]
    \centering
    \begin{subfigure}[b]{0.475\textwidth}
        \centering
        \includegraphics[width=\textwidth]{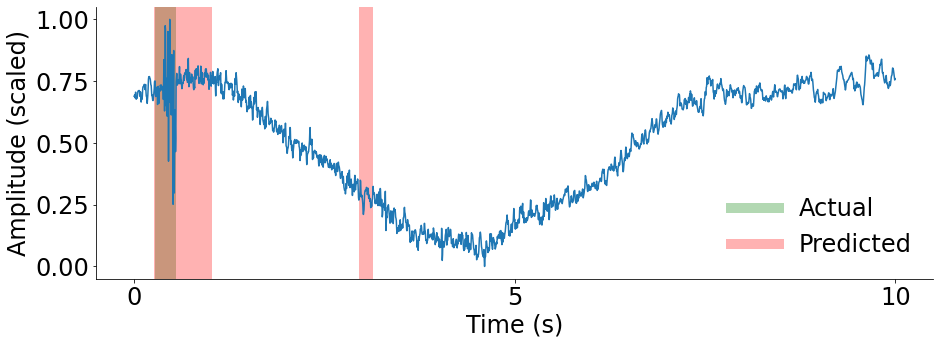}
        \caption[]%
        {{\small Chewing, exit 1}}    
        \label{fig:chewing exit 1}
    \end{subfigure}
    \hfill
    \begin{subfigure}[b]{0.475\textwidth}  
        \centering 
        \includegraphics[width=\textwidth]{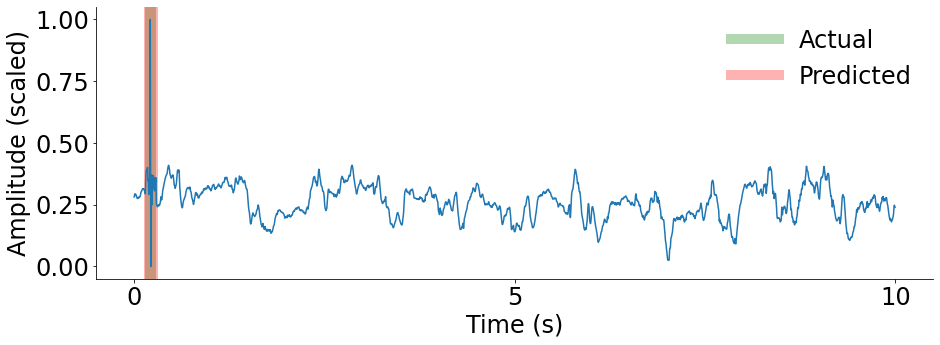}
        \caption[]%
        {{\small Electrode, exit 1}}    
        \label{fig:electrode exit 1}
    \end{subfigure}
    \vskip\baselineskip
    \begin{subfigure}[b]{0.475\textwidth}   
        \centering 
        \includegraphics[width=\textwidth]{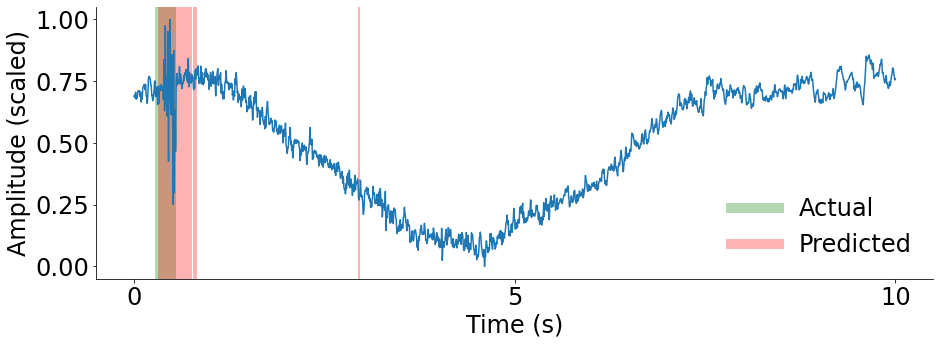} 
        \caption[]%
        {{\small Chewing, exit 5}}    
        \label{fig:chewing exit 4}
    \end{subfigure}
    \hfill
    \begin{subfigure}[b]{0.475\textwidth}   
        \centering 
        \includegraphics[width=\textwidth]{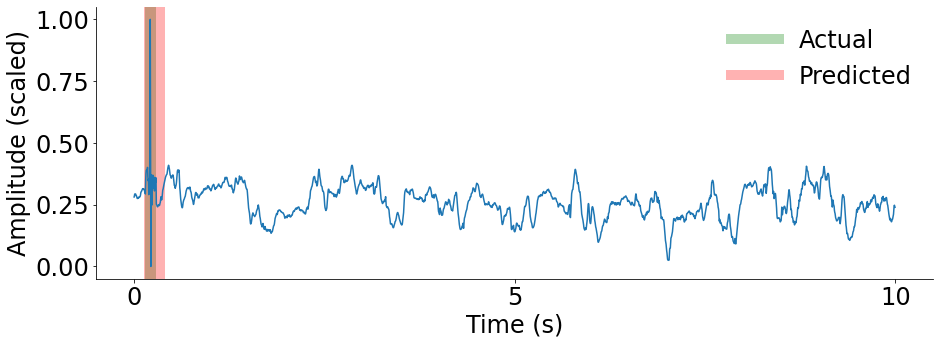}
        \caption[]%
        {{\small Electrode, exit 5}}    
        \label{fig:electrode exit 4}
    \end{subfigure}
    \caption[]{\small Per time point \e artifact predictions for Exits 1 and 5 within a 10 second window. Figure (a) and (c) show disagreement on how the artifact is distributed across time. Figure (b) and (d) show exit disagreement on the end point of the artifact. 
    
    [Green = Actual, Red = Predicted, Brown = True prediction (Red + Green)]}
    \label{fig:chewing and electrode exits}
\end{figure*}

Table~\ref{artifact} shows example results of per exit artifact predictions on the test dataset. The first two artifacts refer to Figure~\ref{fig:chewing and muscle movement exits} whilst the second two rows refer to Figure~\ref{fig:chewing and electrode exits}. 

\begin{table}[hb]
\centering
\resizebox{0.49\textwidth}{!}{
\begin{tabular}{@{}llllll@{}}
\toprule
\multirow{2}{*}{\textbf{Artifact}} & \multicolumn{5}{c}{\textbf{F1 Score $\uparrow$}}                    \\ \cmidrule(l){2-6} 
                          & \textbf{Exit 1} & \textbf{Exit 2} & \textbf{Exit 3} & \textbf{Exit 4} & \textbf{Exit 5} \\ \midrule
Eye movement                      & 0.60   & 0.71   & 0.73   & 0.73   & 0.74   \\
Muscle movement                      & 0.99   & 0.96   & 0.92   & 0.92   & 0.94   \\
Electrode                      & 0.92   & 0.78   & 0.84   & 0.84   & 0.84   \\
Chewing                      & 0.70   & 0.71   & 0.79   & 0.81   & 0.79   \\ \bottomrule
\end{tabular}}
\caption{Average per time point F1 scores of artifact predictions across each exit. Highly disagreeing exits have a larger difference in F1 score. }
\label{artifact}
\end{table}

Estimating uncertainty helps in isolating cases where the model is guessing at random as we can see from the predictions further from the occurrence of the true chewing artifact in Figure~\ref{fig:chewing exit 1} and~\ref{fig:chewing exit 4}. Figure~\ref{fig:electrode exit 1} and~\ref{fig:electrode exit 4} show an electrode artifact where the model is more accurate at exit 1 (0.92) compared to later exits (0.84<) demonstrating that later exits in the ensemble are not always better at making correct predictions. Exit 4 predicts a longer time frame for the artifact occurrence reflecting greater uncertainty. 

\newpage
\hspace{0.5cm}

\clearpage

\end{appendices}

\end{document}